\documentclass[traditabstract]{aa} 

\usepackage{graphicx}
\usepackage{txfonts}
\usepackage{url}
\usepackage{natbib}
\bibpunct{(}{)}{;}{a}{}{,} 
\usepackage[utf8]{inputenc}
\usepackage{subfigure}
\usepackage{verbatim} 
\begin{document}

   \title{The spectrum of kink-like oscillations of solar photospheric magnetic elements}
   \author{M. Stangalini$^{1}$, F. Berrilli$^{2}$, G. Consolini$^{3}$}
   \institute{$^{1}$ INAF-Osservatorio Astronomico di Roma, 00040 Monte Porzio Catone (RM), Italy\\
   $^{2}$ Universita degli Studi di Roma Tor Vergata, 00133 Roma, Italy \\   
   $^{3}$ INAF-Istituto di Astrofisica e Planetologia Spaziali, 00133 Roma, Italy\\
   \email{marco.stangalini@inaf.it}}

  \abstract 
{Recently, the availability of new high-spatial and -temporal resolution observations of the solar photosphere has allowed the study of the oscillations in small magnetic elements. Small magnetic elements have been found to host a rich variety of oscillations detectable as intensity, longitudinal or transverse velocity fluctuations which have been interpreted as MHD waves. Small magnetic elements, at or below the current spatial resolution achieved by modern solar telescopes, are though to play a relevant role in the energy budget of the upper layers of the Sun's atmosphere, as they are found to cover a significant fraction of the solar photosphere. Unfortunately, the limited temporal length and/or cadence of the data sets, or the presence of seeing-induced effects have prevented, so far, the estimation of the power spectra of kink-like oscillations in small magnetic elements with good accuracy. Motivated by this, we studied kink-like oscillations in small magnetic elements, by exploiting very long duration and high-cadence data acquired with the Solar Optical Telescope on board the Hinode satellite. In this work we present the results of this analysis, by studying the power spectral density of kink-like oscillations on a statistical basis. We found that small magnetic elements exhibit a large number of spectral features in the range $1-12$ mHz. More interestingly, most of these spectral features are not shared among magnetic elements but represent a unique signature of each magnetic element itself. \\}  

   \authorrunning{M. Stangalini}
	\titlerunning{The spectrum of kink oscillations}
\maketitle

\section{Introduction}
Solar surface magnetism manifest over a wide range of scales \citep{2006RPPh...69..563S}, from sunspots down to the small elements of the order of $130-150$ km \citep{2010ApJ...723L.164L} or groups of them at larger scales ($\simeq 1000$ km). At the smallest and intermediate scales, they are detectable either as spectropolarimetric signals or  bright points in \textit{G-band} intensity images \citep[][to mention a few]{2004A&A...422L..63W, 2006SoPh..237...13M, 2010ApJ...723..787V, 2013A&A...554A..65U}, although the magnetic field is not a necessary condition for the appearance of a bright point \citep{2001ApJ...553..449B}.
The recent availability of high-cadence and -spatial resolution data of the solar atmosphere has given access to the detection of MHD waves in small magnetic elements in the solar atmosphere. These studies have been mainly motivated by the necessity of assessing the role of magnetic elements in the framework of the chromospheric and coronal heating debate. It has been estimated, indeed, that small magnetic elements with diameters comparable to or below the present resolving power of the current solar telescopes ($ \sim 100-150$ km) cover a significant fraction ($\simeq 1 \%$) of the solar photosphere \citep{2012A&A...539A...6B}, thus small scale magnetism is though to play a significant role in the energy budget of the upper layers \citep{1981A&A....98..155S} . 
Several works have also addressed the diffusion and the advection of small magnetic elements in the solar photosphere by means of turbulent plasma flows \citep[see for example][]{2011A&A...531L...9M, 2011ApJ...743..133A, 2041-8205-759-1-L17, 0004-637X-752-1-48, 2013ApJ...770L..36G}.\\
It is likely that small magnetic elements may  represent a promising mean through which the energy can be transferred from the photosphere to the upper layers of the Sun's atmosphere. This point has been raised many times at first by means of theoretical considerations and numerical simulations. From the theoretical point of view it has been shown that small magnetic elements can support a large variety of waves, from magnetoacoustic waves to kink, sausage and torsional modes \citep[to name a few]{1978SoPh...56....5R, 1981A&A....98..155S, Edwin1983, Roberts1983, Musielak1989, 1998ApJ...495..468S, Hasan2003, Musielak2003a}. This scenario has been also confirmed by means of state-of-the-art MHD simulations accounting for more realistic and complex physical conditions \citep{lrsp-2005-3, Khomenko2008, Fedun2011, 2012ApJ...755...18V, 2012A&A...538A..79N}.
In particular, \citet{lrsp-2005-3} have reviewed in detail the excitation of waves in a magnetic cylinder. They have demonstrated that kink waves and longitudinal compressive waves may coexist in the same flux tube, these being excited by the horizontal motion of the flux tube itself. \citet{Hasan2003} and \citet{2008ApJ...680.1542H} argued that the photospheric forcing of magnetic elements can generate enough energy, in the form of high frequency waves ($\nu > 10$ mHz), sufficient to heat the magnetized chromosphere.  To this regard, \citet{Musielak2003a} and \citet{Musielak2003} have argued that many MHD modes can be excited at the same time in small magnetic elements, at the hands of the photospheric granular buffeting. Through non linear interactions they indeed demonstrated that kink oscillations induced by buffeting can be easily converted into longitudinal waves inside the flux tube itself, that should be then detectable as Doppler velocity and intensity fluctuations. It has been shown that the rapid footpoint motion due to the turbulent granular buffeting can effectively excite kink waves who can propagate upward and eventually couple with longitudinal waves \citep{1997ApJ...486L.145K, Hasan2003}. \\
On the other hand, despite the extensive theoretical and numerical work on the wave excitation and propagation in small magnetic elements, only more recently these aspects have begun to be commensurably addressed from the observational point of view \citep{1995A&A...304L...1V, 2009Jess,2011ApJ...730L..37M, 2012ApJ...744L...5J, 2012ApJ...746..183J, 2013A&A...549A.116J, 2013A&A...554A.115S}.  More in detail,  \citet{2012ApJ...746..183J} have  detected slow upward propagating longitudinal waves in small magnetic elements, using high cadence broad band 2D data, visible as periodic intensity fluctuations in the range $110-600$ s. In addition, \citet{2013A&A...554A.115S}, by exploiting the high-spatial and temporal resolution provided by Sunrise/IMaX, have reported the signature of the interaction between the horizontal oscillations of the flux tubes and the longitudinal oscillations detected as Doppler velocity oscillations within the flux tubes themselves, which manifest as a fixed phase lag between the two kind of oscillations. By using the longest lived magnetic features in the data, the same authors have also estimated the mean power spectrum of both kink-like and longitudinal velocity oscillations in small magnetic features. They mainly found that kink-like oscillations have a comparatively broader spectrum of oscillations with respect to longitudinal waves, with power up to $\nu \sim 8-10$ mHz, i.e. periods of $120-100$ s. Unfortunately, only a few magnetic elements were found to live enough to allow a good frequency resolution. The goal of this work is to extend these results by exploiting the excellent statistics provided by seeing-free Hinode/SOT data. Our primary aim is therefore to study the power spectrum of kink-like oscillations in magnetic elements in the solar photosphere with high frequency resolution.  Since very small scale magnetic elements are found to live for a very limited amount of time (a few minutes, \citet{2006SoPh..237...13M}), we restrict our attention to longer-lived magnetic structures with larger diameters ($800-1000$ km). Similar magnetic structures has been found to host Alfv{\'e}n waves, which are also relevant to the chromospheric heating debate \citep{2009Jess}.

\section{Data set and analysis}
\begin{figure}[t]
   \centering
   \subfigure {\includegraphics[width=10cm, clip]{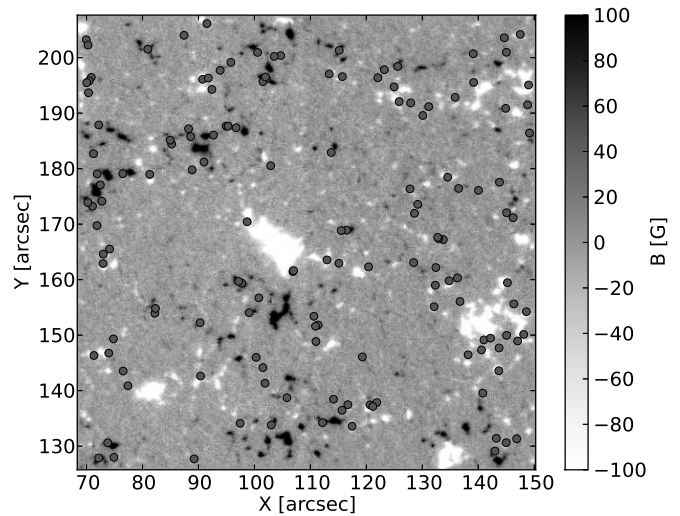}}
   \caption{Time averaged magnetogram obtained by averaging over the whole four-hours data set (saturated between $-100$ G and $100$ G). The red dots indicate the mean position of the longest-lived magnetic elements considered in the analysis.}
    \label{map}
   \end{figure}

The data set used in this work consists of a sequence of high spatial resolution magnetograms acquired by SOT/NFI, the narrow band imager on board Hinode satellite \citep{springerlink:10.1007/s11207-008-9174-z}. The magnetograms were obtained in the Na I $589.6$ nm spectral line from shutterless V and I Stokes filtergrams taken on 2008 August 18, close to the disc centre. The cadence of the data is $30$ s. The diffraction limit in NFI filtergrams at $589.6$ nm is $0.24$ arcsec, while the pixel scale is set at $0.16$ arcsec. This results in a slight spatial downsampling in the focal plane, providing a $2$-pixel Nyquist spatial sampling of $0.32$ arcsec (or $\sim  230$ km on the solar photosphere).\\
The field of view (FoV) is approximately $80 \times 80$ arcsec, and the total duration of the time sequence is approximately $4$ hours. In Fig. \ref{map} the average magnetogram is shown. The FoV encompasses a few supergranules whose borders are highlighted by the network patches visible in the figure.\\
Along with the standard SOT/FG calibration procedure  (the IDL code \textit{prep-fg} available in the Hinode Solarsoft package), we also applied an additional calibration to limit the effects of jitter and tracking residuals. The visual inspection of the data have in fact revealed a low frequency trend in the tracking and a few sudden shifts of the FoV which were not properly handled by the calibration code. Since we are mainly interested in the study of transverse oscillations of small magnetic elements, it is important to properly co-align the data set with subpixel accuracy. A registration procedure allowing subpixel accuracy must be applied. This procedure is based on FFT cross-correlation and utilizes the whole FoV to estimate the misalignment between two images. We applied iteratively the FFT registration, until the mean residuals were minimized. This happened in three registration iterations.\\
To study the dynamics of small magnetic features we applied the YAFTA tracking algorithm \citep{Welsch2003, 2007ApJ...666..576D}. This algorithm identifies and tracks magnetic pixels belonging to the same local maximum. To ensure the reliability of the results, three constraints were applied. Each magnetic feature tracked must lie above a threshold and must have an area slightly larger than the full width at half maximum (FWHM) of the PSF (a box $4 \times 4$ pixels in our case) and a lifetime long enough to ensure a high spectral resolution. The threshold on the magnetic signal is chosen to be $2 \sigma$. Following \citet{2012SoPh..279..295L} we estimated the sigma of the magnetic signal by fitting a Gaussian to the low-field pixels with absolute value of the magnetic flux density below $200$ G. This resulted in a sigma  of $11.8$ G. Lastly, only the magnetic elements with a lifetime greater than $1800$ s are considered in the analysis. Among the whole sample of elements tracked we restricted our attention to a narrow subset of elements with  diameter in the range  $800-1100$ km. This is done to select only similar magnetic elements.
      \begin{figure}[!ht]
   \centering
   \subfigure[] {\includegraphics[width=6cm, clip]{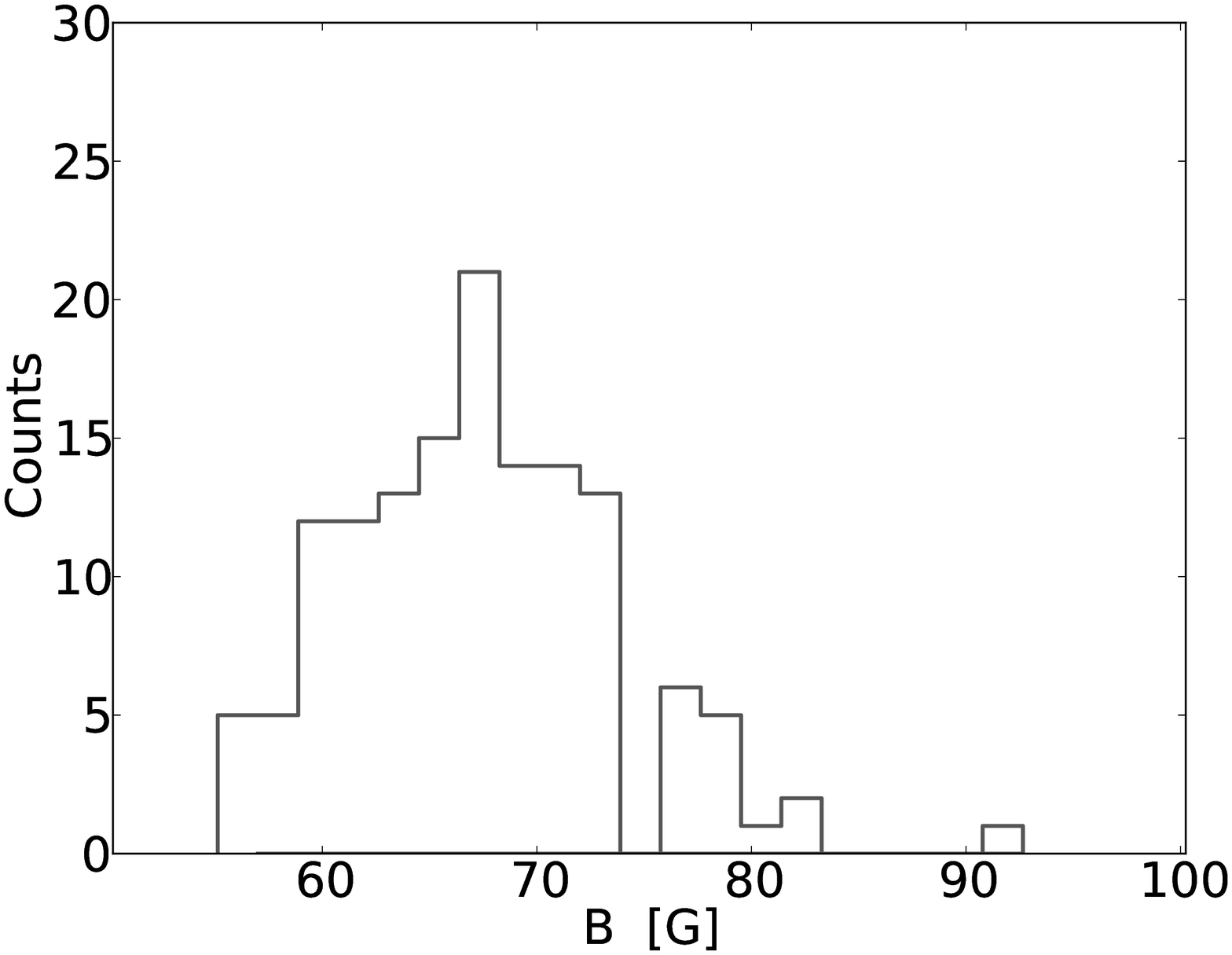}}
   \subfigure[] {\includegraphics[width=6cm, clip]{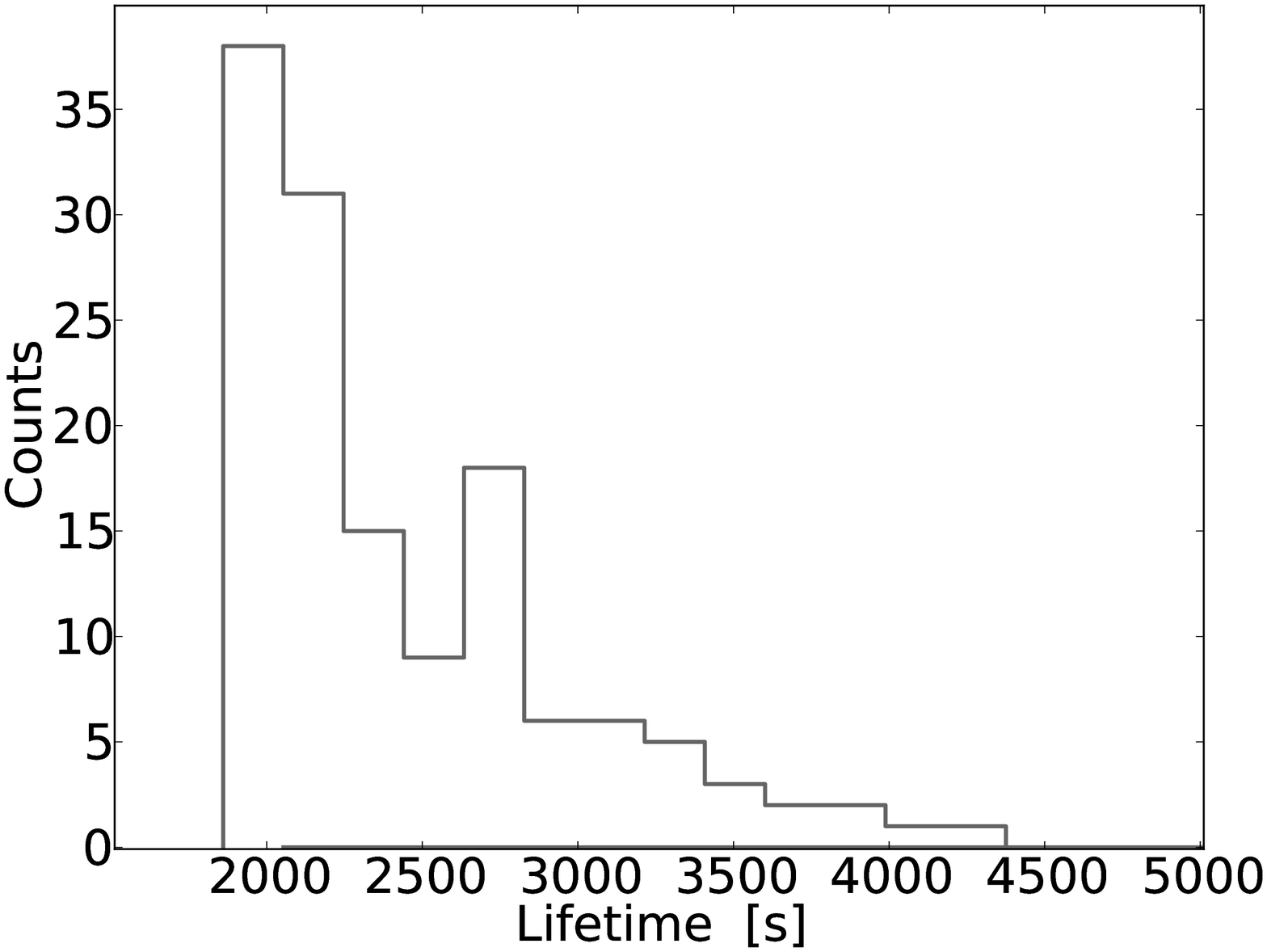}}
   \subfigure[] {\includegraphics[width=6cm, clip]{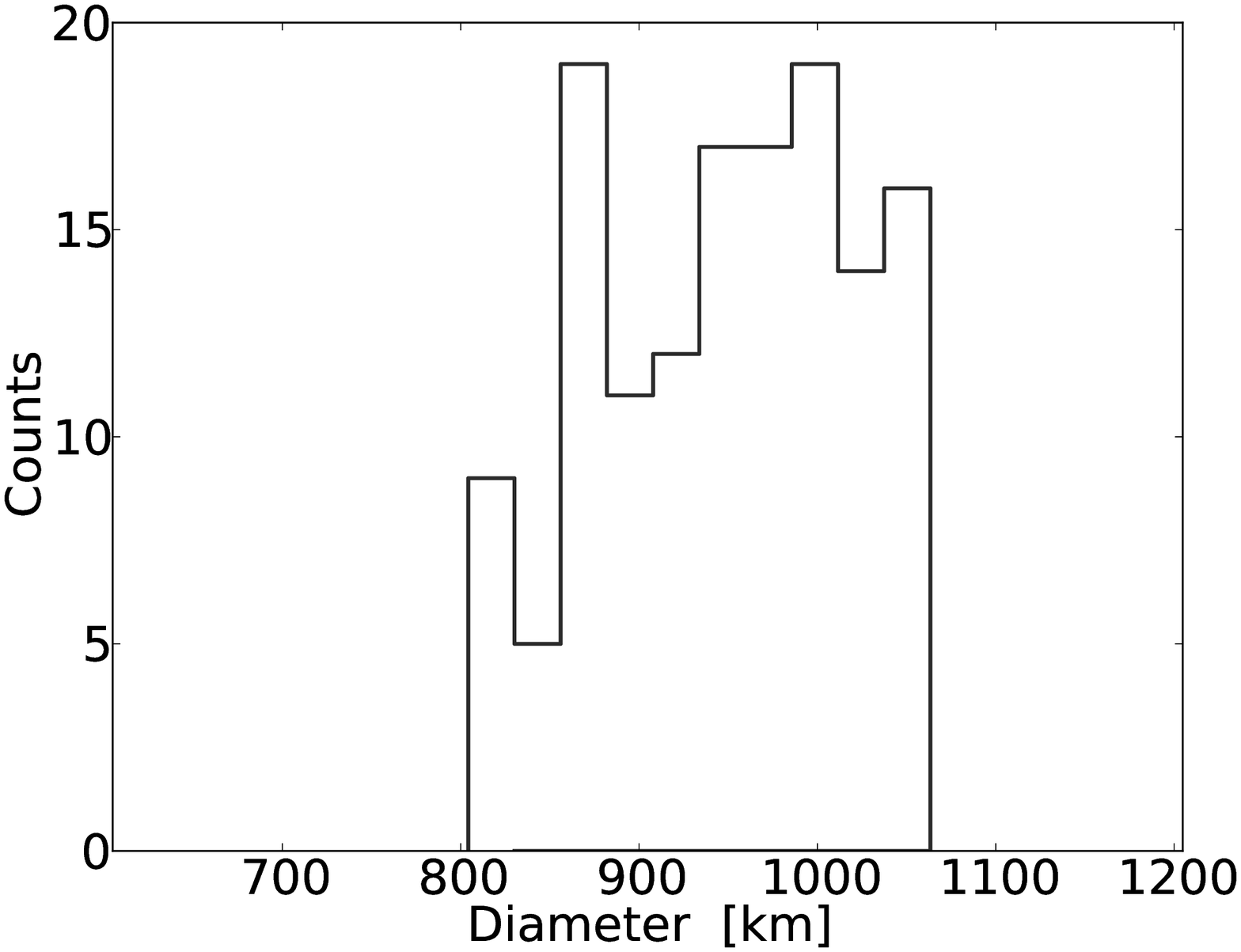}}

   \caption{(a) Histogram of $B$ for the collected magnetic elements. (b) Histogram of the lifetimes of the selected magnetic elements.(c) Histogram of the equivalent diameter of the selected elements.}
    \label{hist}
   \end{figure}  
It is worth noting that the $\sigma$ chosen does not represent the instrumental noise, therefore its choice is completely arbitrary. Our main goal is to limit the detection of spurious magnetic signals close or below the noise level. This is indeed also ensured by the use of the area threshold and and by restricting the analysis to those magnetic features living for more than a given temporal limit; the latter will be further discussed later in the following sections. \citet{2011SoPh..269...13T, 2013A&A...554A.115S} have extensively tested this tracking code on high resolution data.\\
In Fig. \ref{map} (panel b) we show the average position of each magnetic element tracked. We only plot magnetic features  with a minimum lifetime equal to $1800$ s which are those considered in the analysis described in the following sections. The total number of magnetic elements fulfilling the above criteria amount to $146$.\\
From the position of each magnetic element given by the tracking algorithm the horizontal velocities were estimated. The position of the magnetic elements is estimated with subpixel accuracy as a flux-weighted average of the coordinates of the pixels constituting the magnetic element itself. The computation of the horizontal velocity using both the $x$ and $y$ components may suffer from a frequency doubling if the components themselves are oscillating around zero. This is a direct consequence of the definition of the total velocity in which the squared value of the components is considered. In order to avoid this drawback in the spectral analysis we considered only the $x-$component of the horizontal velocity, hereafter $v_{h}$ for simplicity.

\section{Results}
\subsection{Statistical distributions of the sample of magnetic elements}
Before analysing the spectral properties of $v_{h}$, in this section we report the characterizing statistical features of the sample of magnetic elements collected. As aforementioned, we restrict our attention to only those features living more than $1800$ s, this being the limit imposed on the further analysis, discussed in the following section. This limit is set in such a way that a high frequency resolution is preserved in the spectral analysis. \\
In Fig. \ref{hist} we show the statistical distributions of the spatially averaged magnetic field $B$ and of the lifetime of the magnetic elements as obtained from the tracking. In Fig. \ref{hist} the distribution of $|B|$ (panel a) is shown. It is computed as the average value over the area of each magnetic element. From this distribution it is clear that most of the magnetic elements have values in the range $60-75$ G.  In panel (b) of the same figure, we show the histogram of lifetimes which are limited by a threshold at $1800$ s put to ensure a good frequency resolution over the whole sample of magnetic elements, for spectral analysis purposes. In panel (c) we also show the distribution of the equivalent diameter which is defined assuming the magnetic elements to be circular. As mentioned above this parameter has been used to restrict the range of values, in order to select similar magnetic elements in term of physical conditions.\\
Rather than being representative of the magnetic population in the photosphere, these distributions should be better considered as representative of the sampled population of magnetic features, with the aim of explicitly providing its characterization in terms of dynamical properties. It is worth stressing, in fact, that all the distributions indicate the presence of a very small class of sampled magnetic features. In particular the average magnetic field is limited to a very narrow range of values indicating similar magnetic elements in terms of physical parameters.

   \begin{figure*}[t]
   \centering
   \subfigure []{\includegraphics[width=6.cm]{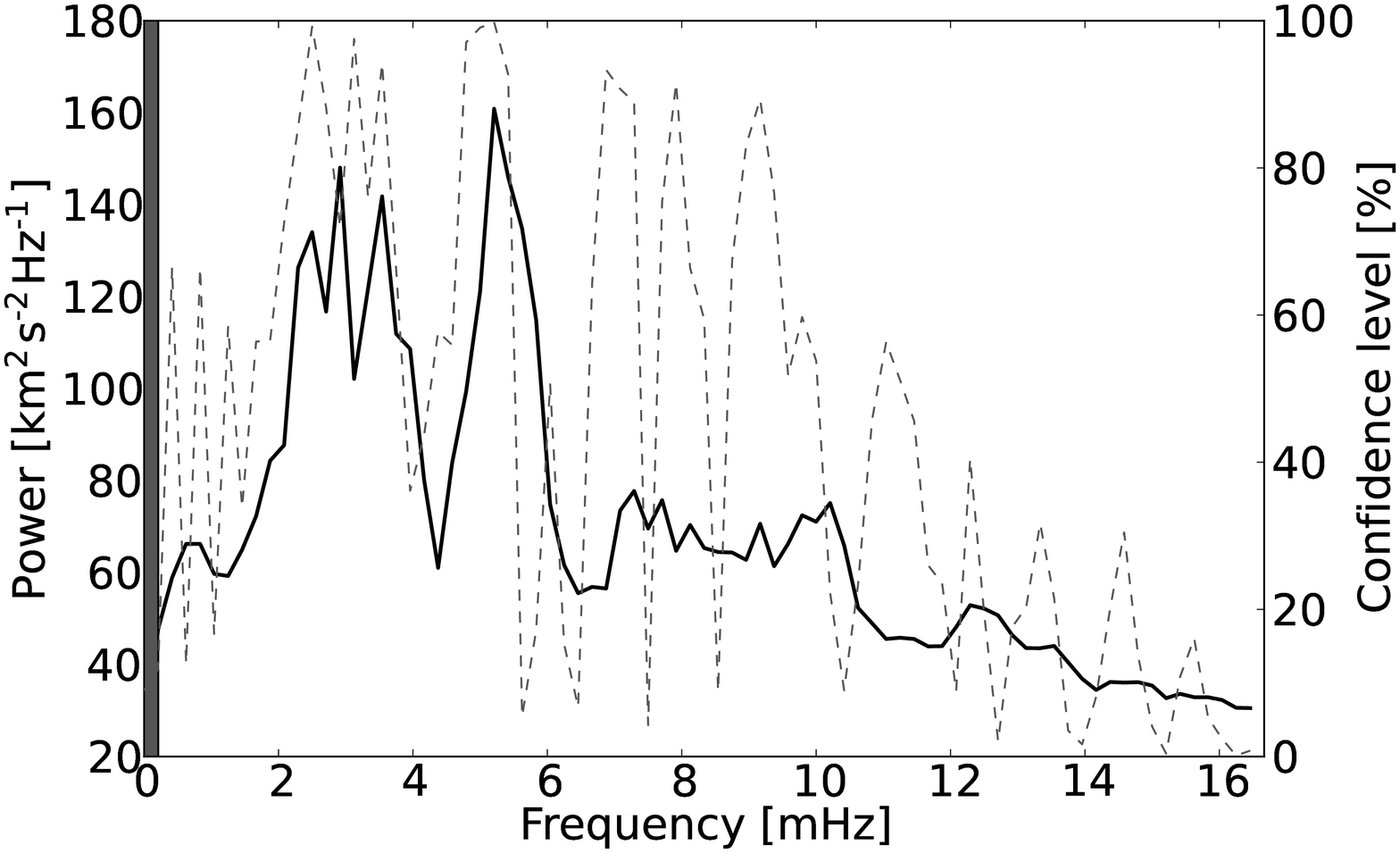}}
   \subfigure []{\includegraphics[width=6.cm]{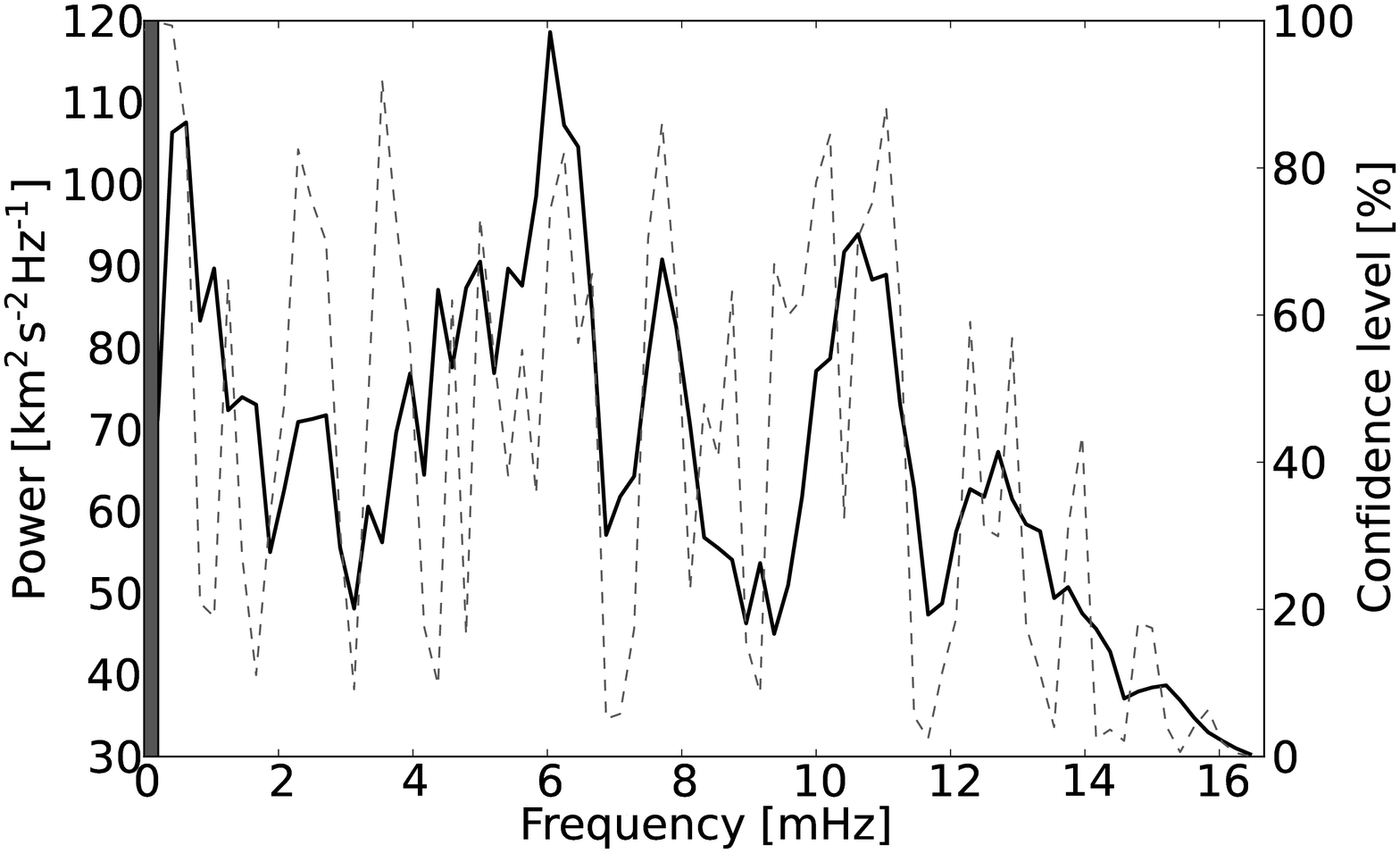}}
   \subfigure []{\includegraphics[width=6.cm]{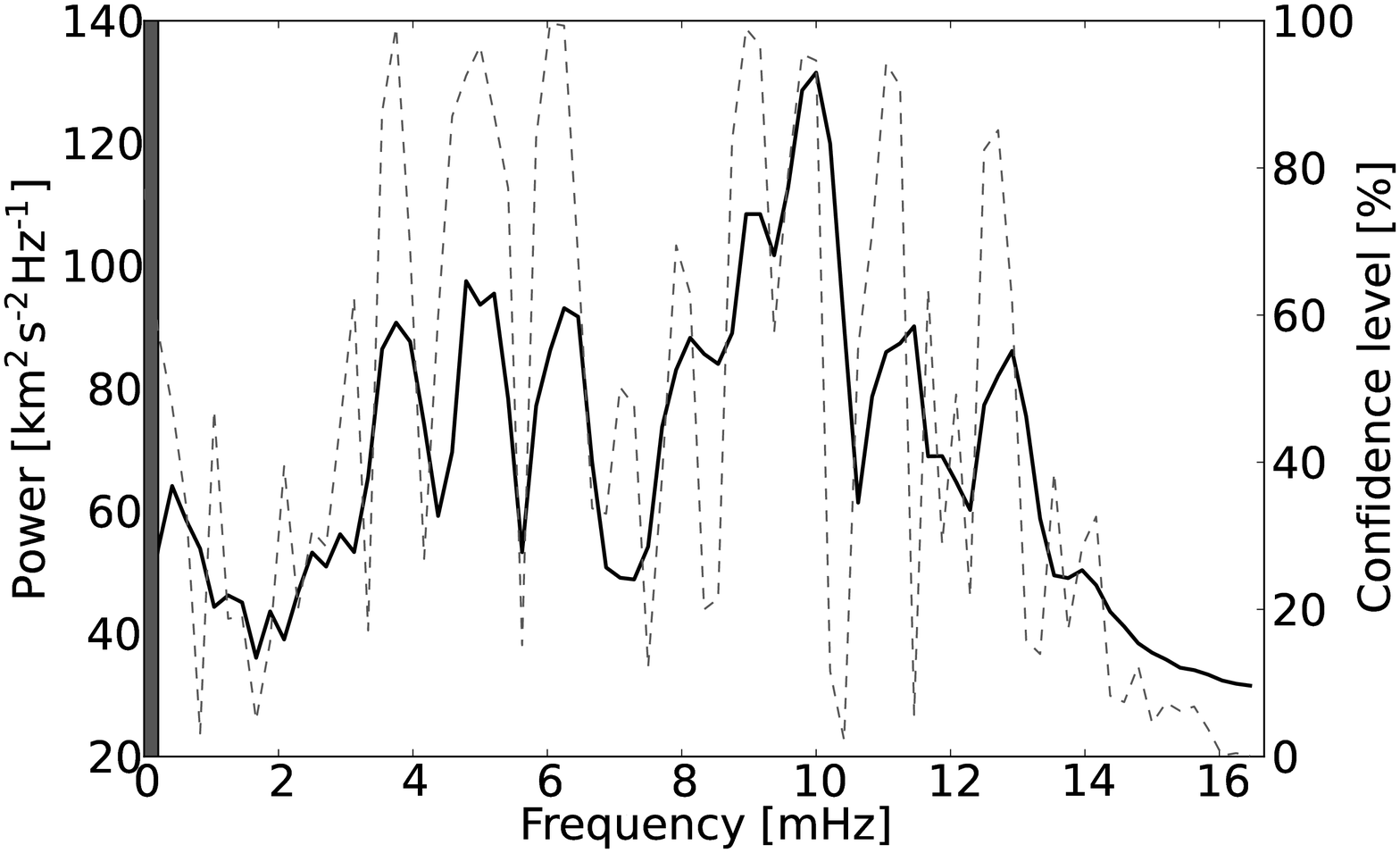}}\\
   \subfigure[]{\includegraphics[width=6.cm]{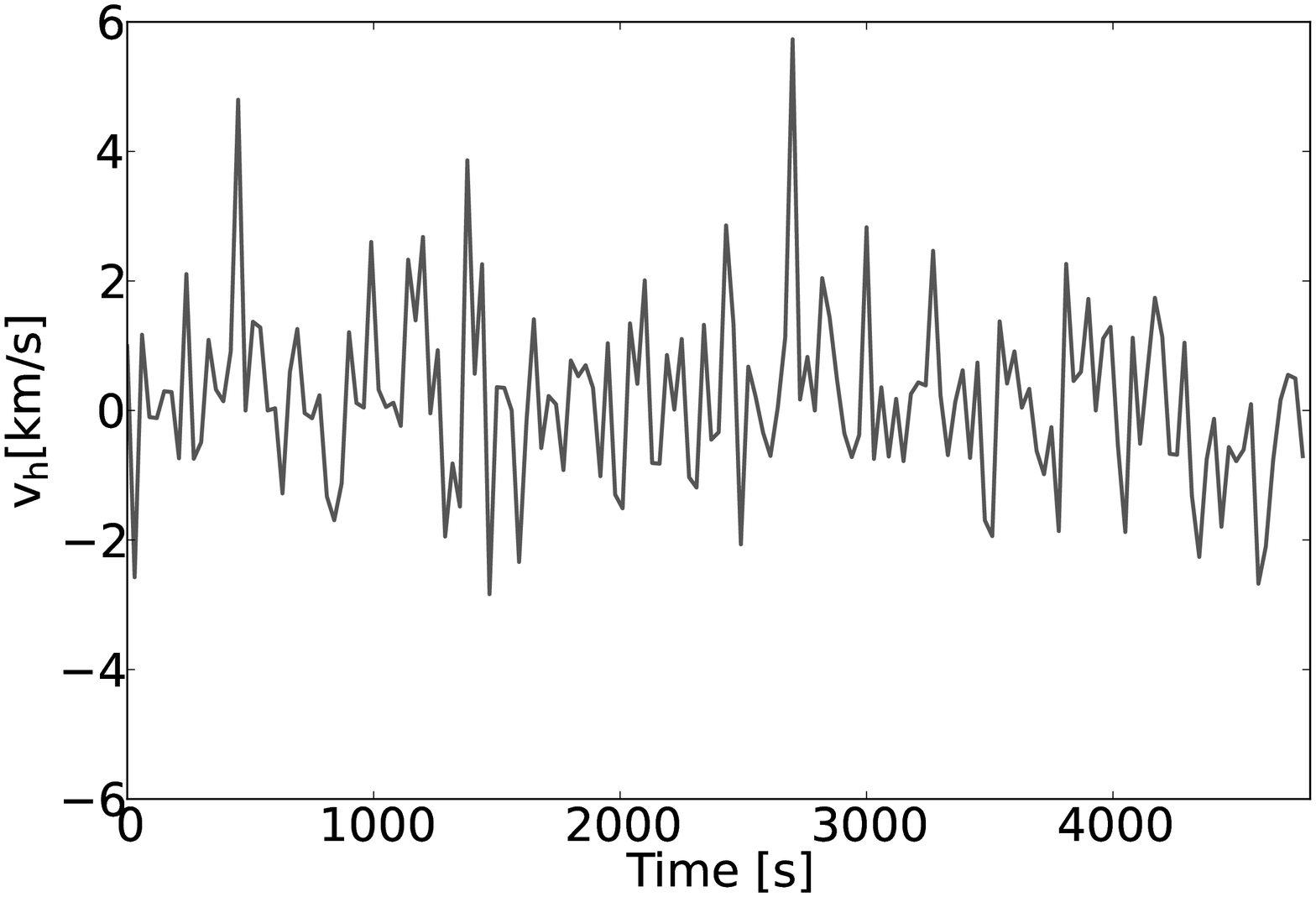}}
   \subfigure[]{\includegraphics[width=6.cm]{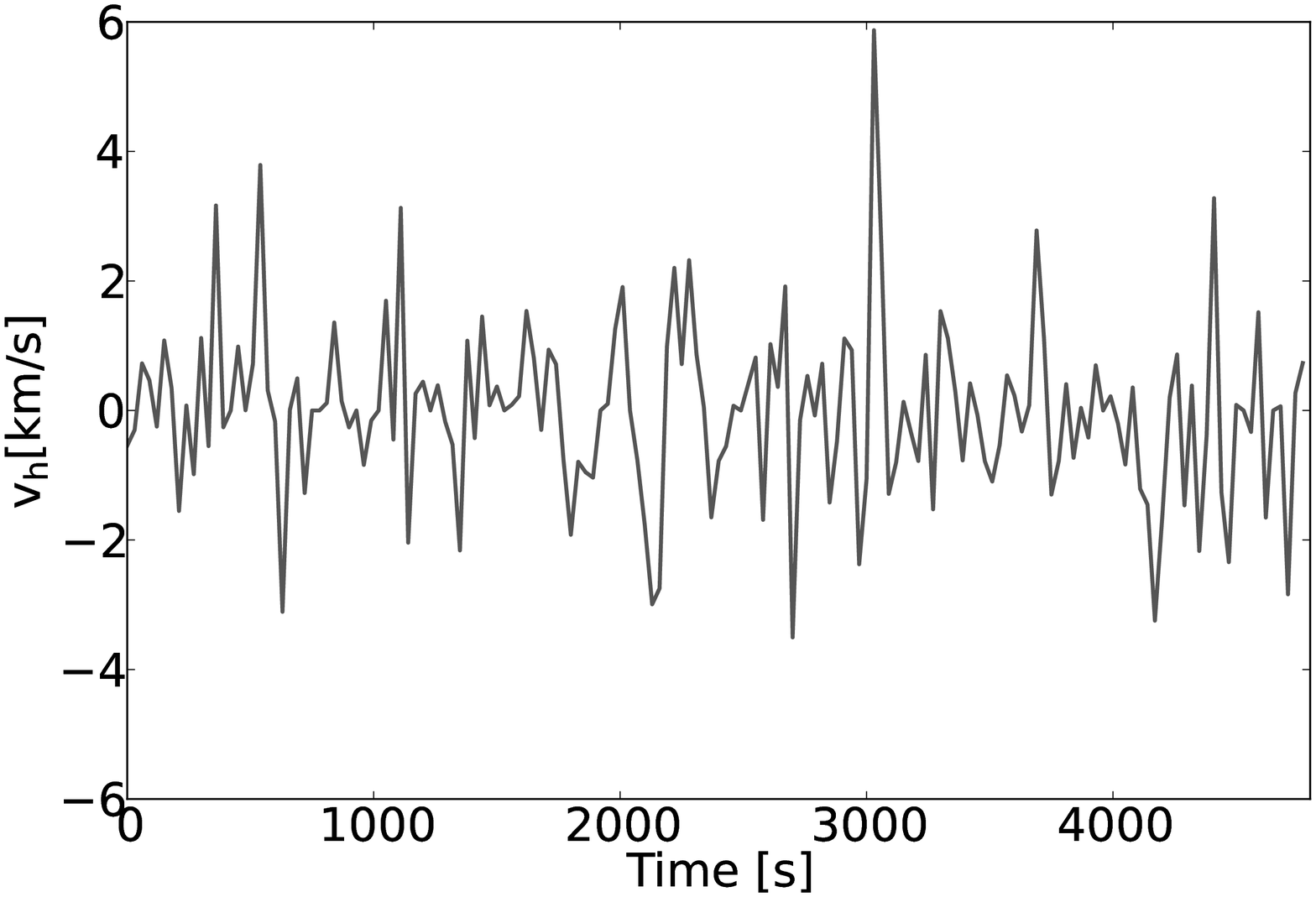}}
   \subfigure[]{\includegraphics[width=6.cm]{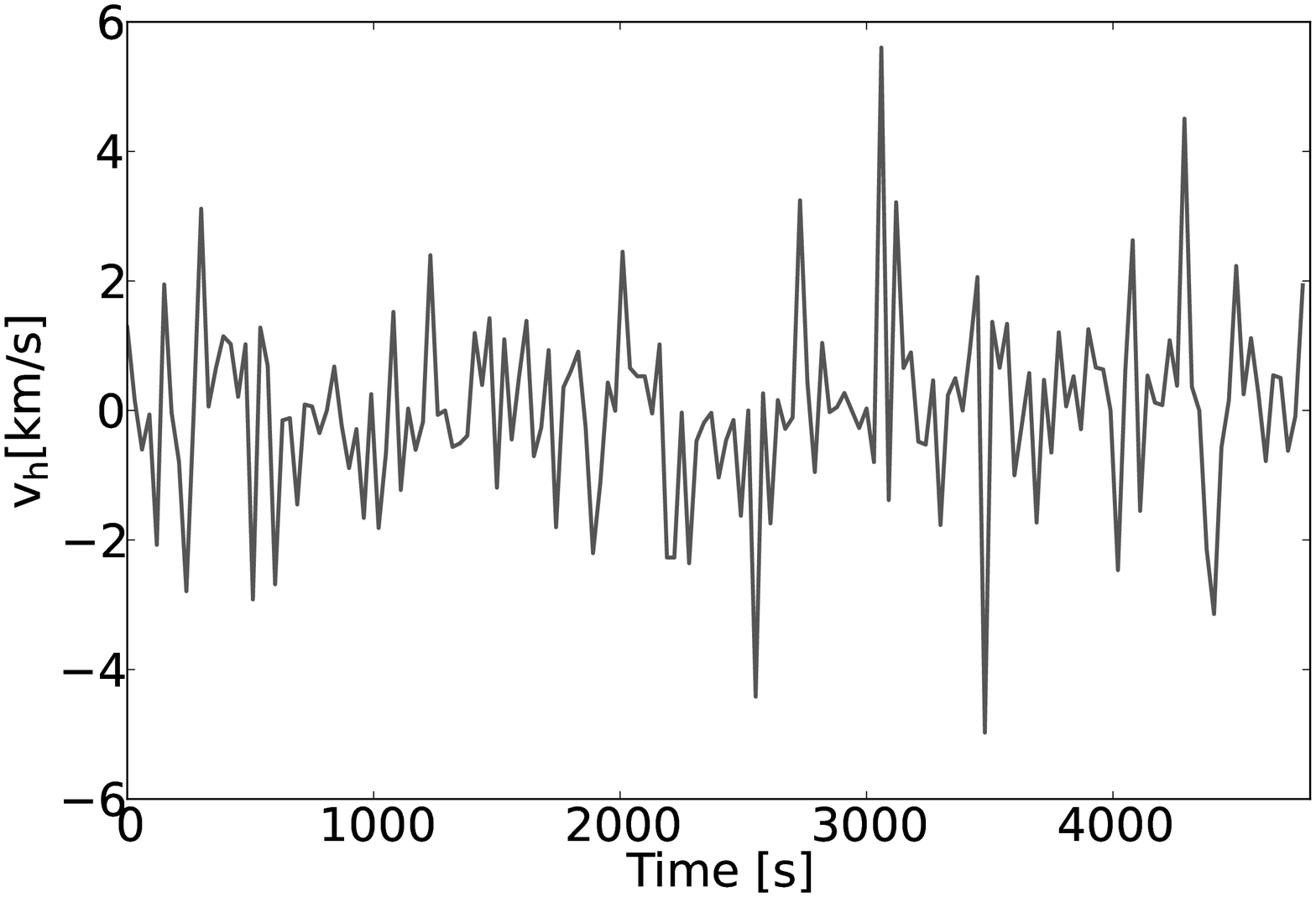}}\\
   \caption{Upper panels: three examples of power spectral densities (continuous line) associated to three typical magnetic elements chosen among the longest lived ones. The dashed line represents the confidence level obtained through a randomization test (see right hand axis of the same plots).The red area represents the frequency resolution imposed by the duration of the time series associated.} Bottom panels: horizontal velocity relative to the power spectra shown.  
    \label{spectra}
   \end{figure*} 

\subsection{Spectral analysis}
It is well known that the sample periodogram estimated from either the FFT or the autocovariance function (ACF) is a biased estimator of the true power spectral density \citep[see for example][]{1982ApJ...263..835S}.
To estimate the periodogram of $v_{h}$, corresponding to the kink-like oscillations of the tracked flux tubes, we followed the commonly accepted procedure described in \citet[Chapter $7$]{chatfield2003analysis} and listed below:

\begin{enumerate}
  \item Prewhitening: the time series $v_{h}$ is detrended with a first order function.
  \item Windowing: $v_{h}$ is then windowed by using a Hanning window.
  \item Periodogram estimation: the FFT of the sample ACF of $v_{h}$ is computed.
  \item The periodogram so obtained is then smoothed by averaging its ordinates in sets of $m$ values. The size of the bins is commonly chosen to be $m=N/40$, where $N$ is the total length of the time series.
\end{enumerate}

At least three effects must be carefully taken into account when estimating the power spectral density of a time series: the variance of the periodogram, the spectral leakage and the time series truncation. While the truncation can be mitigated by multiplying the sequence by a window function with the aim of smoothly reducing the effect of the edges of the sequence, which constitutes a substantial source of spectral noise, the variance of the periodogram is in turn probably the most important problem. The variance of each spectral bin of the periodogram does not decrease as the length of the sequence increases, indeed. This is because when the length of the time series increases the spectral resolution increases too, so that the spectral sampling is not preserved. To this regard, one possibility is to average the ordinates of the power spectral density. This is done, unfortunately, to the detriment of the spectral resolution and a trade off must be chosen as in point $4$ \citep[Chapter $7$]{chatfield2003analysis}. \\
The goal of the procedure listed above is therefore to limit these effects and to provide a robust and consistent estimate of the power spectral density. For a detailed analysis of the biases of the power spectral density and the countermeasures that must be adopted, we refer the reader to  \citet[chapter $7$ and references therein]{chatfield2003analysis}.\\ 
In Fig. \ref{spectra} we show the power spectral densities associated at $v_ {h}$ of three of the longest lived magnetic features (upper row) and their respective time series (bottom row). In the same plots we also show the confidence level obtained through a randomization test (dashed line) which was performed to estimate the reliability of the results. The higher the confidence level, the higher the probability that the corresponding peak in the power spectrum is not due to noise. The normalization test is obtained by shuffling the time series $1500$ times. At each permutation the periodogram is computed again. The number of times a peak in the power spectrum is smaller than what found in the real time series gives the probability that a certain peak is due to noise. A similar analysis has been already performed in \citet{oshea01}. The examples shown in Fig \ref{spectra} are not peculiar among the whole set of magnetic features collected, but represent a few typical examples of the spectral features of the sampled class of elements.\\ 
The time series of the horizontal velocity show oscillations with amplitudes between $1-2$ km/s and, occasionally, short duration transients which can reach larger amplitudes up to $6$ km/s. We recall here that the first order trends have been removed from the original time series, by previously fitting a polynomial.\\
The associated periodograms (upper panels) show a large number of spectral features and the presence of multiple peaks similar to harmonics, although their spacing is not always a multiple of a fundamental frequency. Most of these features have a high confidence level above $75 \%$. These spectral features are found over a broad range of frequencies, from low frequencies close to the resolution limit (indicated by the red area), up to $\sim 12$ mHz, with confidence levels above $80 \%$.\\ 
More interestingly, the spectral features appear to be a distinct signature of the physical conditions of each magnetic element.  In the power spectrum of panel \textit{a}, for instance, most of the energy is contained in the band $1-5$ mHz and little energy is contained at higher frequency. In contrast, in the case plotted in panel \textit{b}, most of the energy is instead contained at larger frequencies ($\nu > 5$ mHz), and no strong counterpart of the peaks at $1-5$ mHz of case of panel \textit{a} is found. The spectral features of different magnetic elements are unique and appear to be distinctive of each magnetic element.\\
This fact can be better seen in Fig. \ref{median} where we plot the median power spectrum obtained from more than $400$ cases (panel a). The spectral density is obtained by considering only those time series with at least $1800$ s of total duration. In order to maintain the same frequency resolution, longer time series are truncated at $1800$ s. For each of these time series, the periodogram is computed and from them the median value is estimated in each frequency bin. In this case we did not smooth the frequency axis of each periodogram to reduce the variance as explained before, as the variance reduction is instead achieved by mediating over a large number of cases. \\
From the plot it is clear that the whole set of magnetic features sampled does not share the same spectral features. In this would have been the case, in fact, a common spectral feature would remain after mediating a large number of power spectra associated to the $v_{h}$ of different magnetic elements. Conversely, the resulting averaged power spectrum appears rather smooth, and there is not any apparent spectral feature.\\
The error bars in the plot represents the standard error obtained from the median absolute deviation.  In the same plot we also show the power spectrum obtained from a randomization test (dashed line). In contrast with the case shown in Fig. \ref{spectra}, where the randomization test is performed by randomly shuffling the time series data $1500$ times and estimating from that the confidence level in terms of probability, in this case we only shuffled the data once for each magnetic element and then we considered the median value computed over all the randomized spectra. This randomization is therefore not used to estimate the confidence level at each spectral position, but gives instead the idea of what would have been a spectrum of random fluctuations.

\subsection{Test of reliability of the tracking algorithm}
\begin{figure*}[t]
   \centering
   \subfigure {\includegraphics[width=8cm, clip]{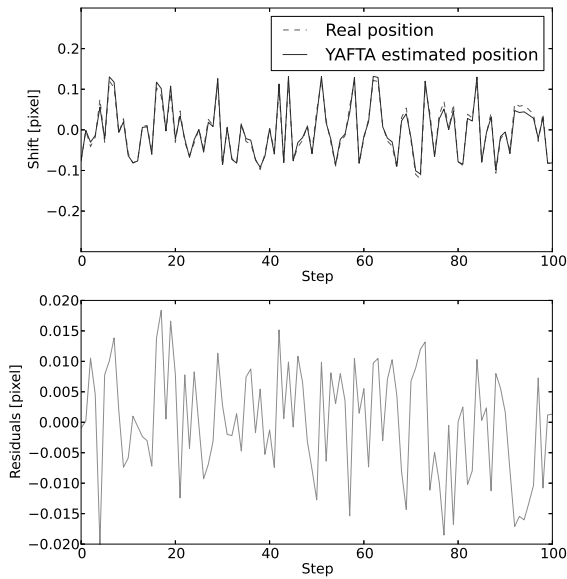}}
   \caption{Test of the tracking algorithm to detect very small shifts of the magnetic elements. Upper panel: The real position of a magnetic element is compared to the one estimated from the tracking algorithm. Lower panel: residuals. }
    \label{test}
   \end{figure*}

In the previous section we have shown that the power spectra of individual magnetic elements are characterized by high frequency features ($12-13$ mHz) with very high confidence level. Of course the ability of detecting such a high frequency oscillations depends on the ability to detect very small shifts in the given time interval ($30$ s in our case). Since the velocity of the magnetic elements is of the order of $1$ km/s and the period associated to the Nyquist frequency (the highest frequency detectable with the provided cadence of the data) is $60$ s, we can estimate the amplitude of oscillation of a magnetic element as $30$ km in the high frequency range. This displacement is much smaller than the pixel scale ($\sim 120$ km). However, the position of each single magnetic element is estimated from the tracking algorithm as a flux-weighted average of the coordinates of each pixel constituting the magnetic element itself, thus with subpixel accuracy. To test the reliability of these estimates we tested the tracking code with simulated data. A syntetic magnetic element with gaussian shape ($FWHM=4$ pixels) is shifted many times by a known amount ( a fraction of pixel) using an FFT technique, and its position is then estimated again through the tracking algorithm. The shifts are uniformly distributed between $-0.12$  and $0.12$ pixels, corresponding to the minimum distance a magnetic element with $1$ km/s of velocity must pass through in $30$ s (the temporal cadence of the data set). If fig. \ref{test} we compare the real position of the magnetic element with the one estimated by the tracking algorithm (upper panel). In the same figure we also plot the residuals between the original position and the estimated one (lower panel). As clear the tracking algorithm is able to detect very small shifts of the magnetic element. This proves that the high frequency features shown in the power spectra of individual magnetic elements (fig. \ref{spectra}) are reliable and their amplitudes are well in the range of capabilities of the tracking algorithm.

\subsection{Estimation of the response function of the flux tubes to the granular buffeting}
\citet{2010ApJ...716L..19M} have estimated the power spectrum of the horizontal velocity in the solar photosphere by using a local correlation tracking technique on long \textit{G}-band sequences of the solar photospheric granulation. They have found that the power spectrum of the horizontal granular velocity shows a double power-law shape with a break at $\simeq 4.7$ mHz. This result has been also introduced in numerical simulation to study the energy budget of MHD waves driven by the photospheric convection, and demonstrating the feasibility of coronal heating through these kind of mechanisms \citep{2010ApJ...710.1857M}. In Fig. \ref{median} (panel b) we also plot in log-log scales the resulting spectrum obtained from \citet{2010ApJ...716L..19M}, for comparison with our estimated power spectrum of kink-like oscillations in small magnetic elements. The two spectra clearly differ in shape with the average spectrum obtained from the tracking of magnetic elements hosting much pore power at high frequencies and less in the low frequency band, compared with the result by \citet{2010ApJ...716L..19M}.\\

It is worth pointing out that when looking at power spectral densities, what is seen is the convolution of the forcing term by the intrinsic response function of the element itself. Thus the observed oscillatory power can be written as follows:

\begin{equation}
P_{obs}(\nu) = F(\nu) \otimes  R(\nu, \alpha_{i}),
\end{equation}

where the symbol '$\otimes$' denotes the convolution in the Fourier space, $F(\nu)$ represents the frequency dependent forcing term, and $R(\nu)$ the response function of the magnetic element depending on the frequency and on the physical conditions  of the flux tube, expressed for simplicity by the set of variables $\alpha_{i}$. Using the forcing spectrum given by \citet{2010ApJ...716L..19M} we can therefore estimate the mean response function of the flux tubes to this forcing. Assuming the average spectrum discussed in the previous section to be the convolution of the forcing term with the average response function of the flux tubes, we can estimate the forcing term as a simple deconvolution in the Fourier space using eq. $1$:

\begin{equation}
R(\nu, \alpha_{i})= P_{obs}(\nu) \otimes F(\nu)^{-1}.  
\end{equation}
The estimated response function deconvolved is also shown in Fig. \ref{median} (panel b). This is characterized by a suppression of power at low frequencies ($\nu < 4$ mHz), and an amplification of the power at shorter periods. This implies that the energy of the photospheric velocity field is best converted into kink oscillations at high frequencies.\\

\section{Discussions and conclusions}
\begin{figure*}[!ht]
   \centering
       \subfigure{\includegraphics[width=8cm, clip]{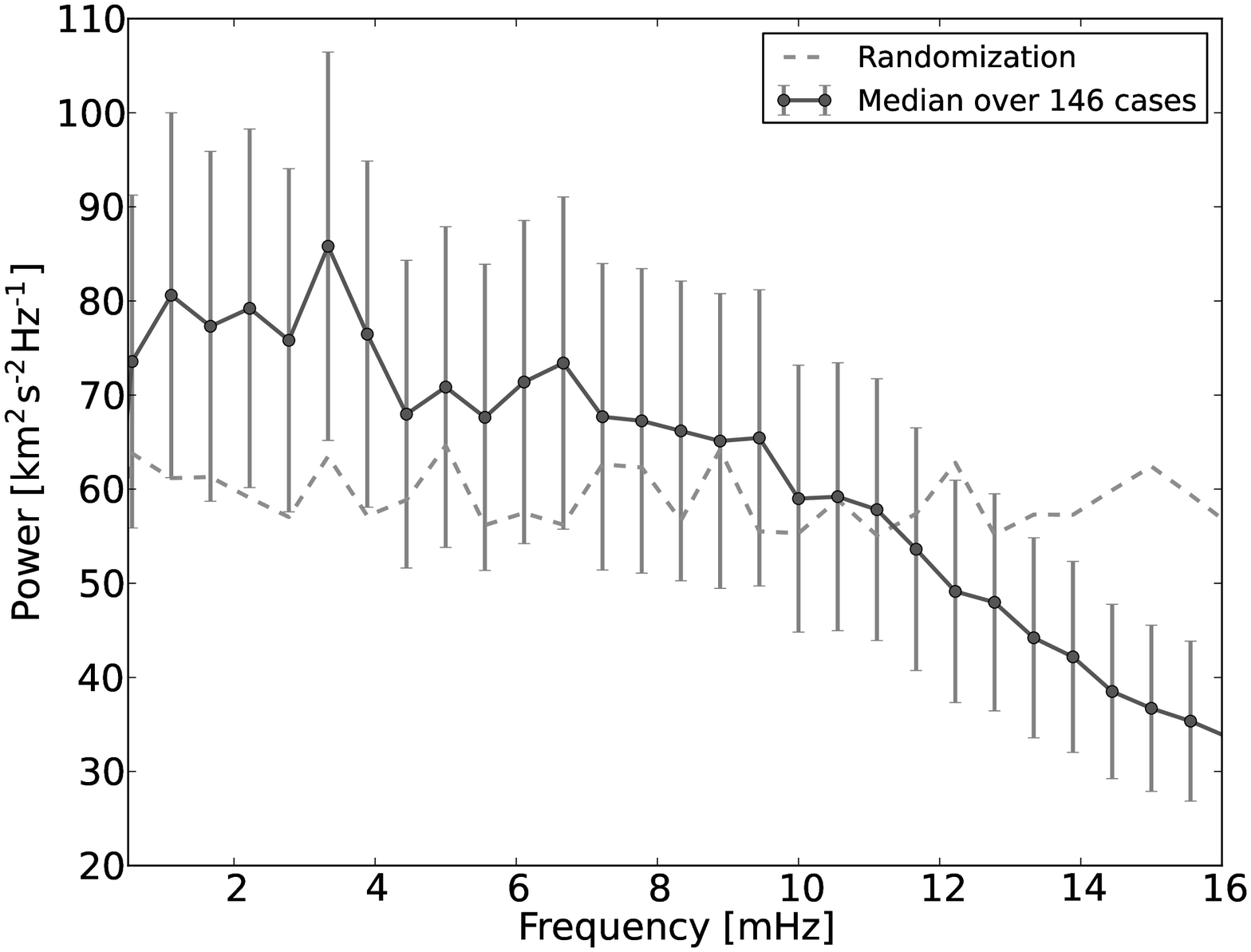}}
    \subfigure {\includegraphics[width=8cm, clip]{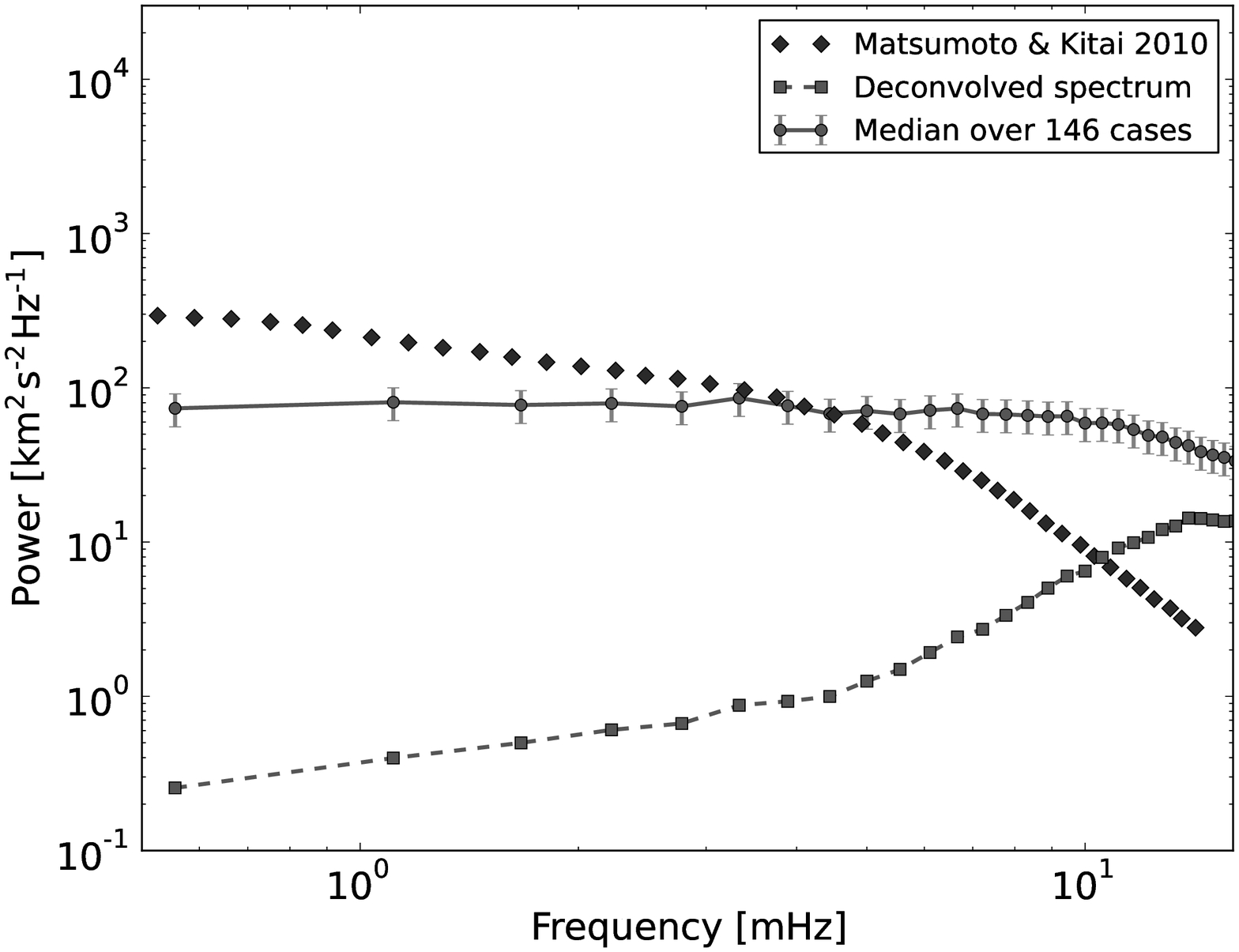}}

   \caption{Upper panel: median power spectrum obtained from $147$ cases (blue), and randomization spectrum (dashed line). Lower panel: same spectrum in log-log axes, buffeting spectrum obtained from \citet{2010ApJ...716L..19M} (green), and deconvolved spectrum (red).  } 
    \label{median}
   \end{figure*}

In this work, by exploiting the fast cadence and the long duration of the SOT/Hinode data at our disposal, we have studied the spectrum of kink-like oscillations in small magnetic elements in the solar photosphere, with very high frequency resolution. Our results shows  many distinct spectral features over a wide range of frequencies; from low frequencies ($1$ mHz) up to $ \sim 12$ mHz. In addition, we have found that occasionally $v_{h}$ shows large transients with amplitudes up to $5-6$ km/s.  By performing a randomization test, it is found that most of these spectral features of the power spectra have a high confidence level ($> 80 \%$) even in the high-frequency regime ($\nu > 8 mHz$).\\
More interestingly, our results also demonstrate that each single magnetic element exhibits its own spectral signature composed by a series of peaks which are not commonly shared among the whole set of magnetic elements tracked. This result can be more easily inferred by looking at Fig. \ref{median}, where we plot the median power spectral density obtained by mediating over more than $140$ magnetic elements. In this plot there is no clear evidence of isolated spectral features that would have appeared, instead, if any common periodicity was present in most of the magnetic elements. This means that the observed  oscillations depend on the magnetic element itself and the spectral features can be considered as the signature of each flux tube and its physical conditions.  We can therefore reasonably assume that the strong differences seen in the periodograms of single magnetic elements are due to the intrinsic response of  each of them to the external plasma forcing. However, as shown in Sect. $3.1$, the class of sampled magnetic elements appears rather restricted to a narrow spectrum of physical conditions expressed by the limited spread of magnetic field values and diameters. This restricts the number of possibilities to explain the variation of the spectral signatures observed since the magnetic flux is very similar for all the cases (e.g. the restoring force provided by the magnetic tension is comparable). Since it is unlikely that the external forcing is different for each magnetic element (this would imply a totally different convection regime like for example in the proximity of a large sunspot), we can reasonably argue that the cause of the differences observed in the spectral signature (i.e. the response of the flux tube to the external forcing) of each magnetic element are associated to non-local parameters indeed. Some example could be the topological connection of the flux tube and its length. However, the data at our disposal are not ideal to investigate this and from them we can only conclude that the $v_{h}$ spectra of oscillation are determined by non-local physical conditions. This of course does not mean that local parameters such as $B$ and the temperature do not take place in determining the spectral features observed, but simply that the narrow distribution of $B$ associated to our sample does not allow to study the effect of the local physical parameters. \\
 It is worth noting, however, that the median power spectrum does show a small peak at $3.5$ mHz. Although the error bars are not small enough to ensure its significance this may represent a signature of the interaction between longitudinal and transverse waves similar to that found by \citet{2013A&A...554A.115S}. Small scale magnetic elements have been found to host five-minute acoustic-like MHD waves, indeed \citep[see for example ][ and references therein]{2011ApJ...743..142H}. However the median power spectrum does not show the presence of a common high frequency peak like those shown in fig. \ref{spectra} representing the power spectra of three typical cases.  \\
By mediating more than $140$ spectra of different flux tubes, we have estimated the mean spectrum which must be considered, in general, as the convolution of the forcing spectrum and the response of the flux tubes. By solving the convolution problem and assuming the spectrum of the granular turbulent buffeting estimated by \citet{2010ApJ...716L..19M}, we have estimated the mean response of the flux tubes to the external granular forcing. This is shown in Fig. \ref{median}. This analysis has shown that the flux tubes act like a high-pass filter, damping oscillations at low frequency and enhancing the response at high frequency. Very recently \citet{2013ApJ...768...17M} have criticized the results from \citet{2010ApJ...716L..19M}. By using the Rapid Oscillations in the Solar Atmosphere instrument \citep[ROSA, ][]{2010SoPh..261..363J}, they have found indeed no evidence of wave energy decreasing with increasing frequency in small magnetic elements in the solar chromosphere as the results by \citet{2010ApJ...716L..19M} would suggest. The results of \citet{2013ApJ...768...17M} are therefore in good agreement with ours. However, for the reasons explained in Sect.  $3.3$ we believe that the spectrum of the horizontal velocity of photospheric plasma flows found by \citet{2010ApJ...716L..19M} using the Local Correlation Tracking (LCT) may not be easily compared to the spectrum of horizontal oscillations observed in small magnetic elements. The first one should be rather considered as the forcing term while the latter represents the response of the flux tubes to this. By adopting this assumption we have estimated the mean response function of the flux tubes in the Fourier space subject to the granular forcing by solving the convolution problem. This estimate is plotted in Fig. \ref{median} and shows that the flux tubes can amplify their response at high frequencies. We note that the use of a LCT tracking method utilizing a correlation window of fixed size ($0.4"$ in the analysis of \citet{2010ApJ...716L..19M}) may underestimate the power at high frequency since the estimated velocity amplitude is dependent on the window size and affected by the smoothing effect of the window itself. The shortcomings may result in systematic errors up to $\sim 30 \%$ \citep{1999ApJ...511..436S, 1999A&A...348..621S, 2004ApJ...616.1242D, 2007ApJ...657.1157G}. As a result, the increased amplification at high frequencies of the estimated response function may be overestimated.  \\ 
However, indipendently of the supposed validity of this assumption, our results demonstrate that the small magnetic element can host a lot of power at high frequency up to $\sim 12$ mHz. This is evident in the power spectra of single magnetic elements (see Fig. \ref{spectra}) and in the mean power spectrum of Fig. \ref{median} where significant power is still contained in the high frequency band ($\nu > 6-10$ mHz).
It has been shown \citep{1981A&A....98..155S} that the cutoff frequency for transversal waves in a flux tube is always smaller than the acoustic cutoff frequency:

\begin{equation}
\omega_{kink}= \dfrac{\omega_{ac}}{2 \gamma (2 \beta +1)} ,
\end{equation}

where $\omega_{ac}$ represents the acoustic cutoff which is $\simeq 5.3$ mHz, and $\beta$ the gas to magnetic pressure ratio. Under typical conditions the transversal cutoff frequency in a flux tube is at least twice as low as the acoustic cutoff. This value is therefore well above the frequency resolution of our data, in the trusted range of the power spectral density diagram. Similarly to acoustic waves, waves below the cutoff are evanescent, while waves whose frequency is above $\omega_{kink}$ can propagate upward in the Sun's atmosphere. Our results show peaks in the power spectra which are mainly in the propagating regime, thus they may represent a substantial source of energy for the upper layers of the atmosphere, after their conversion to longitudinal modes \citep{1997ApJ...486L.145K}. In addition, as highlighted in \citet{2013ApJ...768...17M}, in the high frequency regime processes like resonant absorption, phase mixing, and io-neutral damping are particularly effective. For this reason the high frequency oscillations found in this work in small magnetic elements in the solar photosphere may represent a central ingredient to the energy transfer from the photosphere to the upper layers of the Sun's atmosphere.   \\

\begin{acknowledgements}
This work has been partly supported by the PRIN-INAF 2010 grant, funded by the Italian National Institute for Astrophysics (INAF). Hinode is a Japanese mission developed and launched by ISAS/JAXA, collaborating with NAOJ as a domestic partner, NASA and STFC (UK) as international partners. Scientific operation of the Hinode mission is conducted by the Hinode science team organized at ISAS/JAXA. This team mainly consists of scientists from institutes in the partner countries. Support for the post-launch operation is provided by JAXA and NAOJ (Japan), STFC (U.K.), NASA, ESA, and NSC (Norway).\\
\end{acknowledgements}

\bibliographystyle{aa}
\bibliography{lib}
\end{document}